\documentclass[12pt]{article}

\usepackage[latin1]{inputenc}
\usepackage{tabularx}
\usepackage[dvips]{graphicx}
\usepackage{amsmath}

\def\appendix#1{
  \addtocounter{section}{1}
 \setcounter{equation}{0}
  \renewcommand{\thesection}{\Alph{section}}
 \section*{Appendix \thesection\protect\indent \parbox[t]{11.715cm} {#1}}
  \addcontentsline{toc}{section}{Appendix \thesection\ \ \ #1}
  }

\begin{document}

\begin{titlepage}

\begin{flushright}
UUITP-12/03\\
hep-th/0306147
\end{flushright}

\vspace{1cm}

\begin{center}
{\huge\bf String bits without doubling}
\end{center}
\vspace{7mm}

\begin{center}

{\large Ulf Danielsson, Fredric Kristiansson,\\[2mm]
Martin L\"ubcke and Konstantin Zarembo\footnote{ Also at
ITEP, B.~Cheremushkinskaya 25, 117259 Moscow, Russia}
}

\vspace{7mm}

Institutionen f\"or Teoretisk Fysik, Box 803, SE-751 08
Uppsala, Sweden

\vspace{7mm}

{\tt
ulf.danielsson@teorfys.uu.se, fredric@teorfys.uu.se,\\
martin.lubcke@teorfys.uu.se, konstantin.zarembo@teorfys.uu.se}

\end{center}

\vspace{5mm}

\begin{center}
{\large \bf Abstract}
\end{center}
\noindent
We construct a string bit model in the pp-wave background
in which fermion doubling produces the correct spectrum
of string states.

\end{titlepage}


\section{Introduction}

The type IIB superstring theory in the pp-wave background with Ramond-Ramond
flux \cite{Metsaev,Metsaev:2002re} has drawn considerable attention
recently, mainly because it arises in a certain limit of AdS/CFT
correspondence \cite{Blau:2001ne,Blau:2002mw, BMN} and describes a
particular set of operators in the dual super-Yang-Mills (SYM) theory \cite
{BMN}. The relationship between string states and field-theory operators, a
cornerstone feature of the gauge/string theory duality \cite{Aharony:1999ti}%
, is remarkably simple and explicit in the pp-wave limit. The operators in
some sense correspond to discretized strings which are built out of a finite
number of partons \cite{BMN}. The parton picture of the string, and also the
fact that string theory in the pp-wave is most easily quantized in the
light-cone gauge, is suggestive of the use of the string bit models which
were proposed a long time ago to describe strings of partons in the
light-cone frame \cite{Giles:1977mp}. The key observation is that the length
of the string in the light-cone gauge is proportional to light-cone momentum 
$P^{+}$ \cite{Goddard:qh}. Therefore it can be interpreted
as an ordered array of
partons, each of which carries a fixed small portion of the $P^{+}$. The
continuous
string is recovered in the limit when the number of partons becomes
infinite. 

In the context of the pp-wave string partons are elementary fields of the
SYM theory and the string Hamiltonian arises from their perturbative
interactions \cite{BMN}. The string bit model in the pp-wave background was
introduced in \cite{Verlinde,Vaman:2002ka} and gives an elegant unified
description of the multi-string Hilbert space and of the string interactions.
The string interactions correspond to $1/N$ corrections in the SYM theory,
and it was proposed that the string bit model can describe operator mixing
in the SYM theory non-perturbatively in $1/N$ \cite
{Vaman:2002ka,Pearson:2002zs,zhou} (see \cite{Beisert:2002ff,Freedman:2003bh} for
the discussion of operator mixing on the field-theory side). 

It was noticed in \cite{Bell_Soch} that pp-wave string bit
model potentially suffers from the fermion doubling problem. 
The fermion doubling is not specific to the pp-wave string and was 
discussed earlier for the supersymmetric string in flat space
\cite{BT}. 
The fermion doubling implies
the appearance
of spurious low-energy modes of fermions in lattice field theory
 \cite{lattice}. The wave functions of these 
modes are sign-alternating with the period of two lattice spacings, i.e. they
have a momentum of the order of the lattice cutoff.  Their energy is
nevertheless finite. In the context of the string bit model,
doublers threaten to destroy the correct string spectrum in the continuum
limit and should be eliminated before the continuum limit is taken.
In \cite{BT}, where the flat-space superstring
was discussed, it was proposed to add a mass term for doublers
in such a way that they become infinitely massive in the continuum limit.
It would be interesting to generalize this method to the pp-wave
background.
We advocate an alternative approach to the doubling problem.
Instead of removing the doublers, we propose to reduce the fermion content
of the string bit model by half and to regard doublers as physical states.
Our approach closely follows one of the well-known methods to tackle 
fermion doubling
in lattice field theories \cite{Kogut:ag,Susskind:1976jm}. The
fermion doubling will automatically recover the correct string spectrum
in the continuum limit. 
An important point to note here is that
 doubling produces a pair of identical
fermions, and
indeed there are two world-sheet fermions in the type IIB string theory 
with the same quantum numbers. 

The outline of the paper is as follows. In section two we introduce a new
string-bit Hamiltonian, in section three we propose new supersymmetry
generators, in section four we investigate their algebra, and in section
five, finally, we present some conclusions.

\section{The string-bit Hamiltonian}

The light-cone Hamiltonian for the IIB Green-Schwarz superstring in the
pp-wave background is given by \cite{Metsaev}\footnote{%
We use the notations and conventions of \cite{bible} for $SO(8)$ spinors and
Dirac matrices. The word-sheet fermions are in the $\mathbf{8_{s}}$ of $SO(8)
$. 
The mass matrix $\Pi _{ab}=\gamma _{a\dot{a}}^{1}\gamma _{\dot{a}%
c}^{2}\gamma _{c\dot{b}}^{3}\gamma _{\dot{b}b}^{4}$ is symmetric and
traceless, and explicitly breaks $SO(8)$ to $SO(4)\times SO(4)$.} 
\begin{equation}
H_{\mathrm{continuum}}=\frac{1}{2}\int d\sigma \,\left( p^{2}+x^{\prime
}{}^{2}+\ x^{2}+i\theta \theta ^{\prime }-i\tilde{\theta}\tilde{\theta}%
^{\prime }+2i\theta \Pi \tilde{\theta}\right) .  \label{eq:cont_Hamiltonian}
\end{equation}
The essential ingredient of the string-bit model is a discretized version of
this Hamiltonian. The main advantage of the discretization is the possibility to
describe string interactions in a combinatorial way. 
The Fock space of a discretize string naturally
includes multi-string states, and consequently string interactions can
be introduced by just adding extra terms to the Hamiltonian \cite{Verlinde}. 
Such a simple description of interactions
is impossible for the continuous string. 
Our main goal, however, is to address the doubling problem, which 
arises before the Fock space is extended to include multiple strings, and we
shall concentrate on the Hilbert space of a single string in what follows.

The string bit Hamiltonian of \cite{Verlinde} is the straightforward
discretization of (\ref{eq:cont_Hamiltonian}): 
\begin{eqnarray}
H_{\mathrm{string\,bit}} &=&\frac{a}{2}\sum_{n=1}^{J}\left[
p_{n}^{2}+x_{n}^{2}+\frac{1}{a^{2}}(x_{n+1}-x_{n})^{2}\right.   \notag
\label{hamilton} \\
&&\qquad \qquad \left. +\frac{i}{a}\,\theta _{n}\theta _{n+1}-\frac{i}{a}\,%
\tilde{\theta}_{n}\tilde{\theta}_{n+1}-2i\tilde{\theta}_{n}\Pi \theta _{n}%
\right] .
\end{eqnarray}
To be more precise, this Hamiltonian is the gauge-fixed version of the
string bit model. We believe, however, that a gauge-invariant generalization
of our approach, and an inclusion of multi-string states, should not cause
any serious problems. In the above expression the inverse lattice spacing $%
1/a$ should be identified with the 't~Hooft coupling of the dual SYM theory 
\cite{Verlinde}, and the phase space coordinates are understood to obey the
following canonical commutation relations 
\begin{equation}
\big[p_{n}^{i},x_{m}^{j}\big]=-\frac{i}{a}\,\delta ^{ij}\delta _{nm},\qquad %
\big\{\theta _{n}^{a},\theta _{m}^{b}\big\}=\frac{1}{a}\,\delta ^{ab}\delta
_{nm},\qquad \big\{\tilde{\theta}_{n}^{a},\tilde{\theta}_{m}^{b}\big\}=\frac{%
1}{a}\,\delta ^{ab}\delta _{nm}.
\end{equation}
Here $i=1\ldots 8$ and $a=1\ldots 8$ are indices of the vector and of one of
the spinor representations of $SO(8)$. The other spinor representation will
be denoted by dotted indices. 

A straightforward discretization of the supersymmetry generators 
yields: 
\begin{eqnarray}
Q_{\mathrm{string\,bit}} &=&a\sum_{n=1}^{J}\left[ p_{n}^{i}\gamma ^{i}\theta
_{n}-x_{n}^{i}\gamma ^{i}\Pi \tilde{\theta}_{n}+\frac{1}{a}%
(x_{n+1}^{i}-x_{n}^{i})\gamma ^{i}\theta _{n}\right] ,  \notag \\
\widetilde{Q}_{\mathrm{string\,bit}} &=&a\sum_{n=1}^{J}\left[
p_{n}^{i}\gamma ^{i}\tilde{\theta}_{n}+x_{n}^{i}\gamma ^{i}\Pi \theta _{n}-%
\frac{1}{a}(x_{n+1}^{i}-x_{n}^{i})\gamma ^{i}\tilde{\theta}_{n}\right] .
\label{eq:org_Q}
\end{eqnarray}
The Hamiltonian and the supercharges are diagonalized by Fourier transforms
given by 
\begin{eqnarray}
x_{n}=\frac{1}{\sqrt{J}}\sum_{p=-J/2}^{J/2-1}x_{p}e^{2\pi ipn/J}, &\qquad
&p_{n}=\frac{1}{\sqrt{J}}\sum_{p=-J/2}^{J/2-1}p_{p}e^{2\pi ipn/J}, \\
\theta _{n}=\frac{1}{\sqrt{J}}\sum_{p=-J/2}^{J/2-1}\theta _{p}e^{2\pi ipn/J},
&\qquad &\tilde{\theta}_{n}=\frac{1}{\sqrt{J}}\sum_{p=-J/2}^{J/2-1}\tilde{%
\theta}_{p}e^{2\pi ipn/J},
\end{eqnarray}
which leads to a Hamiltonian of the form 
\begin{eqnarray}
H &=&\frac{a}{2}\sum_{p=-J/2}^{J/2-1}\left[ p_{p}p_{-p}+x_{p}x_{-p}+\frac{4}{%
a^{2}}x_{p}x_{-p}\sin ^{2}\frac{p\pi }{J}\right.   \notag \\
&&\qquad \qquad \left. +\ \frac{1}{a}\,(\theta _{p}\theta _{-p}-\tilde{\theta%
}_{p}\tilde{\theta}_{-p})\sin \frac{2p\pi }{J}-2i\tilde{\theta}_{p}\Pi
\theta _{-p}\right] .  \label{eq:org_H_in_p}
\end{eqnarray}

It is now easy to see that the kinetic energy of fermions has two zeros, at $%
p=0$ and at $p=J/2$, in each Brillouin zone. As a consequence there are
twice as many light fermion states as there should be in the continuum
limit. The bosons do not suffer from this problem, since modes with $p\sim
J/2$ have energies of order of the cutoff and decouple in the continuum
limit. The fermions states with momentum close to $J/2$ are potentially
dangerous, since their energy is finite in the continuum limit and they may
mix with physical states when string interactions are taken into account. In
addition, as shown in \cite{Bell_Soch}, the variation of $H$ under the
supersymmetry transformations fails to vanish.

We propose a mild modification of the string bit model which is free from
the doubling problem. Actually, instead of viewing it as a problem, we shall
take advantage of the fermion doubling. The key observation is that the
spectrum of the type IIB superstring contains two fermions transforming in
the same way. This suggests that the right spectrum can emerge as the result
of the doubling of a single fermion species. The idea is then to start out
with half the number of fermions in the discretized case and let the fermion
doubling provide for the missing states in the continuum limit. 

Since the momenta of the doublers lie close to the boundary of the Brillouin
zone, their wave functions in the coordinate space are approximately sign
alternating with the period of two lattice spacings. We can couple doublers
to the normal low-momentum modes by introducing the sign factor $(-1)^{n}$.
A Hamiltonian that achieves this goal is obtained from (\ref{hamilton}) by
omitting the $\tilde{\theta}$ and introducing the staggered mass term%
\footnote{%
Strictly speaking, we should assume that the number of sites in the lattice
is even. Otherwise, $(-1)^{n}$ is not a periodic function on the lattice,
and the mass term will contain a ''defect'' near $n=0$. Its effect, however,
should disappear in the continuum limit, and for this reason the difference
between lattices with an even and an odd 
number of sites should not affect our main
results. This difference between even and odd lattices
was recognized earlier for the flat-space superstring \cite{BT}.}: 
\begin{eqnarray}
H &=&\frac{a}{2}\sum_{n=1}^{J}\left[ p_{n}^{2}+x_{n}^{2}+\frac{1}{a^{2}}%
(x_{n+1}-x_{n})^{2}\right.   \notag \\
&&\qquad \qquad \left. +\ \frac{i}{a}\,\theta _{n}\theta
_{n+1}-i(-1)^{n}\theta _{n}\Pi \theta _{n+1}\right]  \\
&=&\frac{a}{2}\sum_{p=-J/2}^{J/2-1}\left( p_{p}p_{-p}+x_{p}x_{-p}+\frac{4}{%
a^{2}}\,x_{p}x_{-p}\sin ^{2}\frac{p\pi }{J}\right.   \notag \\
&&\qquad \qquad \left. +\ \frac{1}{a}\,\theta _{p}\theta _{-p}\sin \frac{%
2p\pi }{J}+i\theta _{p+J/2}\Pi \theta _{-p}\cos \frac{2p\pi }{J}\right) .
\end{eqnarray}
In the continuum limit, where $J\rightarrow \infty $ and $~a\rightarrow 0$,
we recover (\ref{eq:cont_Hamiltonian}) 
if we make the identification 
\begin{equation}
\tilde{\theta}_{p}=\theta _{p+J/2}.
\end{equation}
There are two sets of low-energy modes: $\theta _{p}$ and $\tilde{\theta}_{p}
$ with $p\ll J$. In the coordinate representation the physical fermions are
linear combinations of lattice variables on two neighboring sites: 
\begin{eqnarray}
\theta _{n}^{a} &\rightarrow &\frac{1}{2}\big(\theta _{n}^{a}+\theta
_{n+1}^{a}\big)  \notag \\
\tilde{\theta}_{n}^{a} &\rightarrow &\frac{1}{2}(-1)^{n}\big(\theta
_{n}^{a}-\theta _{n+1}^{a}\big).  \label{eq:new_theta}
\end{eqnarray}
That is, $\theta _{p}$ (for finite $p$) corresponds to fluctuations around a
constant $\theta _{n}$, while $\tilde{\theta}_{p}$ corresponds to
fluctuations around a staggered state where $\theta _{n}$ is alternating
between the lattice sites.

\section{The supersymmetry transformations}

We must now turn to the subject of supersymmetry. The discretization of the
supercharges and the definition of the supersymmetry transformations on the
lattice is a very subtle procedure. For instance, the straightforward
substitution of the fermion variables (\ref{eq:new_theta}) into the
supercharges (\ref{eq:org_Q}) does not work. The supersymmetry
transformations then mix low-energy degrees of freedom that survive the
continuum limit with heavy lattice modes, which renders truncation of the
superalgebra to the low-energy sector inconsistent. 
We propose to start with the original supersymmetry generators in (\ref
{eq:org_Q}) above, make replacements according to (\ref{eq:new_theta}), and
in addition replace bosonic variables by their symmetric combinations 
\begin{equation}
p_{n}^{i}\rightarrow \frac{1}{2}\left( p_{n}^{i}+p_{n+1}^{i}\right) ,\qquad
x_{n}^{i}\rightarrow \frac{1}{2}\left( x_{n}^{i}+x_{n+1}^{i}\right) .
\end{equation}
In this way we get for the supercharges: 
\begin{eqnarray}
Q &=&\frac{a}{4}\sum_{n=1}^{J}\left[ \vphantom{\frac{1}{a}}
(p_{n}^{i}+p_{n+1}^{i})\gamma ^{i}(\theta _{n}+\theta _{n+1})\right.   \notag
\\
&&\qquad \quad -\ (-1)^{n}(x_{n}^{i}+x_{n+1}^{i})\gamma ^{i}\Pi (\theta
_{n}-\theta _{n+1})  \notag \\
&&\qquad \quad \left. +\ \frac{1}{a}(x_{n+2}^{i}-x_{n}^{i})\gamma
^{i}(\theta _{n}+\theta _{n+1})\right] ,
\end{eqnarray}
and 
\begin{eqnarray}
\widetilde{Q} &=&\frac{a}{4}\sum_{n=1}^{J}\left[ \vphantom{\frac{1}{a}}%
(-1)^{n}(p_{n}^{i}+p_{n+1}^{i})\gamma ^{i}(\theta _{n}-\theta _{n+1})\right. 
\notag \\
&&\qquad \quad +\ (x_{n}^{i}+x_{n+1}^{i})\gamma ^{i}\Pi (\theta _{n}+\theta
_{n+1})  \notag \\
&&\qquad \quad \left. -\ \frac{1}{a}(-1)^{n}(x_{n+2}^{i}-x_{n}^{i})\gamma
^{i}(\theta _{n}-\theta _{n+1})\right] .
\end{eqnarray}
As we shall now show, these operators have the desired properties.
Straightforward calculations give the following supersymmetry
transformations: 
\begin{eqnarray}
\big[Q,x_{n}^{i}\big] &=&-\frac{i}{4}\gamma ^{i}(\theta _{n+1}+2\theta
_{n}+\theta _{n-1}), \\
{\big[Q,p_{n}^{i}\big]} &=&\frac{i}{4}(-1)^{n}\gamma ^{i}\Pi (\theta
_{n+1}-2\theta _{n}+\theta _{n-1})  \notag \\
&&-\ \frac{i}{4a}\gamma ^{i}(\theta _{n+1}+\theta _{n}-\theta _{n-1}-\theta
_{n-2}), \\
\big\{Q,\theta _{n}\big\} &=&\frac{1}{4}(p_{n+1}^{i}+2p_{n}^{i}+p_{n-1}^{i})%
\gamma ^{i}  \notag \\
&&-\ \frac{1}{4}(-1)^{n}(x_{n+1}^{i}+2x_{n}^{i}+x_{n-1}^{i})\gamma ^{i}\Pi  
\notag \\
&&-\ \frac{1}{4a}(x_{n+2}^{i}+x_{n+1}^{i}-x_{n}^{i}-x_{n-1}^{i})\gamma ^{i},
\end{eqnarray}
and 
\begin{eqnarray}
\big[\widetilde{Q},x_{n}^{i}\big] &=&\frac{i}{4}(-1)^{n}\gamma ^{i}(\theta
_{n+1}-2\theta _{n}+\theta _{n-1}), \\
{\big[\widetilde{Q},p_{n}^{i}\big]} &=&\frac{i}{4}\gamma ^{i}\Pi (\theta
_{n+1}+2\theta _{n}+\theta _{n-1})  \notag \\
&&-\ \frac{i}{4a}(-1)^{n}\gamma ^{i}(\theta _{n+1}-\theta _{n}-\theta
_{n-1}+\theta _{n-2}), \\
\big\{\widetilde{Q},\theta _{n}\big\} &=&\frac{1}{4}%
(-1)^{n}(p_{n+1}^{i}+2p_{n}^{i}+p_{n-1}^{i})\gamma ^{i}  \notag \\
&&+\ \frac{1}{4}(x_{n+1}^{i}+2x_{n}^{i}+x_{n-1}^{i})\gamma ^{i}\Pi   \notag
\\
&&-\ \frac{1}{4a}(-1)^{n}(x_{n+2}^{i}+x_{n+1}^{i}-x_{n}^{i}-x_{n-1}^{i})%
\gamma ^{i}.
\end{eqnarray}
After Fourier transforming these become 
\begin{eqnarray}
\big[Q,x_{p}^{i}\big] &=&-i\cos ^{2}\frac{\pi p}{J}\gamma ^{i}\theta _{p},
\label{Qx} \\
\big[Q,p_{p}^{i}\big] &=&-i\cos ^{2}\frac{\pi p}{J}\gamma ^{i}\Pi \theta
_{p-J/2}+\frac{2}{a}\sin \frac{\pi p}{J}\cos ^{2}\frac{\pi p}{J}e^{-i\pi
p/J}\gamma ^{i}\theta _{p}, \\
\big\{Q,\theta _{p}\big\} &=&\cos ^{2}\frac{\pi p}{J}p_{p}^{i}\gamma
^{i}-\sin ^{2}\frac{\pi p}{J}x_{p+J/2}^{i}\gamma ^{i}\Pi   \notag \\
&&\qquad \qquad \qquad \qquad -\ \frac{2i}{a}\sin \frac{\pi p}{J}\cos ^{2}%
\frac{\pi p}{J}e^{i\pi p/J}x_{p}^{i}\gamma ^{i},  \label{eq:Q_theta}
\end{eqnarray}
and 
\begin{eqnarray}
\big[\widetilde{Q},x_{p}^{i}\big] &=&-i\cos ^{2}\frac{\pi p}{J}\gamma
^{i}\theta _{p+J/2}, \\
\big[\widetilde{Q},p_{p}^{i}\big] &=&i\cos ^{2}\frac{\pi p}{J}\gamma ^{i}\Pi
\theta _{p}-\frac{2}{a}\sin \frac{\pi p}{J}\cos ^{2}\frac{\pi p}{J}e^{-i\pi
p/J}\gamma ^{i}\theta _{p+J/2}, \\
\big\{\widetilde{Q},\theta _{p}\big\} &=&\sin ^{2}\frac{\pi p}{J}%
p_{p+J/2}^{i}\gamma ^{i}+\cos ^{2}\frac{\pi p}{J}x_{p}^{i}\gamma ^{i}\Pi  
\notag \\
&&\qquad \qquad \qquad \qquad +\ \frac{2}{a}\cos \frac{\pi p}{J}\sin ^{2}%
\frac{\pi p}{J}e^{i\pi p/J}x_{p+J/2}^{i}\gamma ^{i}.
\end{eqnarray}
In the continuum limit, focusing on the light modes with momenta close to
zero, the transformations reduce to the expected expressions: 
\begin{equation}
\begin{array}{lcl}
\big[Q,x^{i}\big]=-i\gamma ^{i}\theta , & \qquad  & \big[\widetilde{Q},x^{i}%
\big]=-i\gamma ^{i}\tilde{\theta}, \\[2mm]
\big[Q,p^{i}\big]=-i\gamma ^{i}\Pi \tilde{\theta}-2i\gamma ^{i}\theta
^{\prime }, & \qquad  & \big[\widetilde{Q},p^{i}\big]=i\gamma ^{i}\Pi \theta
+2i\gamma ^{i}\tilde{\theta}^{\prime }, \\[2mm]
\big\{Q,\theta \big\}=p^{i}\gamma ^{i}-2x^{\prime }{}^{i}\gamma ^{i}, & 
\qquad  & \big\{\widetilde{Q},\theta \big\}=-x^{i}\gamma ^{i}\Pi , \\[2mm]
\big\{Q,\tilde{\theta}\big\}=x^{i}\gamma ^{i}\Pi , & \qquad  & \big\{%
\widetilde{Q},\tilde{\theta}\big\}=p^{i}\gamma ^{i}-2x^{\prime }{}^{i}\gamma
^{i}.
\end{array}
\end{equation}
A crucial observation is that there is no undesired mixing between low
energy and high energy states. In particular, the $\cos ^{2}\pi p/J$ factor
in (\ref{Qx}) makes sure, for $p$ close to $J/2$, that there is no mixing of
the low energy mode $\tilde{\theta}_{p-J/2}$ into the supersymmetry
transformation of the high energy mode $x_{p}^{i}$. This is a necessary
property if we want to achieve a consistent supersymmetric truncation to the
low energy sector.

\section{Checking the superalgebra}

The results of the previous section are very encouraging, but 
we also need to check that the pp-wave superalgebra emerges in the continuum
limit. Straightforward calculations show that 
\begin{eqnarray}
\lbrack Q,H] &=&-\frac{i}{8a}\sum_{n=1}^{J}\left(
x_{n+2}^{i}-2x_{n}^{i}+x_{n-2}^{i}\right) \gamma ^{i}\left( \theta
_{n}-\theta _{n-1}\right)   \notag \\
&&+\frac{i}{8}\sum_{n=1}^{J}\left( p_{n+2}^{i}-2p_{n}^{i}+p_{n-2}^{i}\right)
\gamma ^{i}\theta _{n}  \notag \\
&&-\frac{i}{8}\sum_{n=1}^{J}\left( x_{n+2}^{i}-2x_{n}^{i}+x_{n-2}^{i}\right)
\left( -1\right) ^{n}\gamma ^{i}\Pi \theta _{n-1}  \notag \\
&&+\frac{ia}{8}\sum_{n=1}^{J}\left(
x_{n+2}^{i}-2x_{n}^{i}+x_{n-2}^{i}\right) \gamma ^{i}\theta _{n}  \notag \\
&&+\frac{ia}{8}\sum_{n=1}^{J}\left(
p_{n+2}^{i}-2p_{n}^{i}+p_{n-2}^{i}\right) \left( -1\right) ^{n}\gamma
^{i}\Pi \theta _{n},
\end{eqnarray}
which can be Fourier transformed to 
\begin{eqnarray}
\lbrack Q,H] &=&-\frac{1}{2}\sum_{p=-J/2}^{J/2-1}\sin ^{2}\frac{2\pi p}{J}%
\bigg[ia(x_{p}^{i}\gamma ^{i}\theta _{-p}+p_{p}^{i}\gamma ^{i}\Pi \theta
_{J/2-p})  \notag \\
&&\qquad \qquad \qquad \qquad \quad +\ i(p_{p}^{i}\gamma ^{i}\theta
_{-p}+e^{2\pi ip/J}x_{p}^{i}\gamma ^{i}\Pi \theta _{J/2-p})  \notag \\
&&\qquad \qquad \qquad \qquad \quad +\ \frac{2}{a}\sin \frac{\pi p}{J}e^{\pi
ip/J}x_{p}^{i}\gamma ^{i}\theta _{-p}\bigg]
\end{eqnarray}
Clearly, the supercharges do not commute with the Hamiltonian, but it is
easy to see that the commutator either vanish in the continuum limit (as $%
J\rightarrow \infty $), or is zero provided that we put the heavy lattice
modes to zero.

We can also check that the anti-commutators of supercharges agree with the
pp-wave superalgebra \cite{Metsaev} up to terms which decouple in the
continuum limit. We find: 
\begin{eqnarray}
\left\{ Q_{\dot{a}},Q_{\dot{b}}\right\}  &=&2\delta _{\dot{a}\dot{b}}(%
\mathcal{H}+\mathcal{P})+2\left( \gamma ^{i}\Pi \gamma ^{j}\right) _{\dot{a}%
\dot{b}}(\mathcal{R}^{ij}+\mathcal{T}^{ij}), \\
\left\{ \widetilde{Q}_{\dot{a}},\widetilde{Q}_{\dot{b}}\right\}  &=&2\delta _{\dot{a}%
\dot{b}}(\mathcal{H}-\mathcal{P})+2\left( \gamma ^{i}\Pi \gamma ^{j}\right)
_{\dot{a}\dot{b}}(\mathcal{R}^{ij}-\mathcal{T}^{ij}), \\
\left\{ Q_{(\dot{a}},\widetilde{Q}_{\dot{b})}\right\}  &=&2\delta _{\dot{a}\dot{b%
}}\mathcal{V}+2\left( \gamma ^{i}\Pi \gamma ^{j}\right) _{\dot{a}\dot{b}}%
\mathcal{S}^{ij}\,,
\end{eqnarray}
where 
\begin{eqnarray}
\mathcal{H} &=&\frac{a}{32}\sum_{n=1}^{J}\left[ (p_{n+2}+2p_{n+1}+p_{n})^{2}+%
\frac{1}{a^{2}}(x_{n+3}+x_{n+2}-x_{n+1}-x_{n})^{2}\right.   \notag \\
&&\left. +(x_{n+2}+2x_{n+1}+x_{n})^{2}+\frac{2i}{a}\,\theta _{n}(5\theta
_{n+1}+\theta _{n+3})-16i(-1)^{n}\theta _{n}\Pi \theta _{n+1}\right] ,
\label{newH} \\
\mathcal{P} &=&\frac{1}{32}\sum_{n=1}^{J}\left(
\{p_{n},x_{n+3}+3x_{n+2}+2x_{n+1}-2x_{n}-3x_{n-1}-x_{n-2}\}+8i\theta
_{n}\theta _{n+2}\right) , \\
\mathcal{V} &=&\frac{a}{16}\sum_{n=1}^{J}(-1)^{n}\left[
-p_{n}(p_{n+2}-p_{n})-\frac{1}{a^{2}}%
\,x_{n}(x_{n+3}-3x_{n+1})+x_{n}(x_{n+2}-x_{n})\right.   \notag \\
&&\left. -\frac{i}{a}\,\theta _{n}(\theta _{n+3}-\theta _{n+1})\right] , \\
\mathcal{R}^{ij} &=&\frac{1}{16}%
\sum_{n=1}^{J}(-1)^{n}x_{n}^{(i}(x_{n+3}-2x_{n+2}-3x_{n+1}+2x_{n})^{j)}\,, \\
\mathcal{T}^{ij} &=&\frac{a}{16}%
\sum_{n=1}^{J}(-1)^{n}p_{n}^{(i}(x_{n+2}-2x_{n}+x_{n-2})^{j)}\,, \\
\mathcal{S}^{ij} &=&\frac{1}{32}%
\sum_{n=1}^{J}x_{n}^{(i}(x_{n+3}+2x_{n+2}-x_{n+1}-4x_{n}-x_{n-1}+2x_{n-2}+x_{n-3})^{j)}.
\end{eqnarray}

The $\delta _{\dot{a}\dot{b}}$ terms in the anti-commutators of $Q$ and $%
\widetilde{Q}$ have the right structure, the other terms are lattice artifacts
and should go to zero in the continuum limit. The operator $\mathcal{H}$ is
just another discretization of the string Hamiltonian and $\mathcal{P}$ can
be regarded as a lattice counterpart of the momentum operator. One should
note, though, that $\mathcal{H}$ do \textit{not }agree with $H$ when
evaluated on the heavy lattice modes. $\mathcal{H}$ would lead to a
similar doubling problem as before, but now for the bosons, and could
not serve as a new Hamiltonian.
This is expected, since we can only
hope for the supersymmetry algebra to close when applied to the low-energy
states. Indeed the difference $\mathcal{H}-H$ goes to zero in the continuum
limit. 
The same happens to the structures that do not appear in the continuum
superalgebra. The operators $\mathcal{R}^{ij}$ and $\mathcal{T}^{ij}$
contain staggered bosons, which means that they necessarily decouple when we
put heavy modes to zero. The symmetric part of the anti-commutator between $Q
$ and $\widetilde{Q}$ is zero for the continuum pp-wave string\footnote{%
The anti-symmetric part, which we do not discuss here, is proportional to a
linear combination of $SO(4)$ rotation generators \cite{Metsaev}.}. The two
terms in the $\{Q,\widetilde{Q}\}$ indeed vanish in the continuum limit, but for
different reasons: the operator $\mathcal{V}$ is of the staggered boson
type, while $\mathcal{S}^{ij}$ contains a discretized version of the second
derivative of $x$ which lacks one power of the inverse lattice spacing.

\section{Conclusions}

In this paper we have proposed a modification the string bit model \cite
{Verlinde} in which the fermion doubling is taken into account. By taking
advantage of the doubling, we construct two fermions in the continuum from a
single lattice fermion and its doubler. We have shown that the correct
spectrum of the type IIB string in the pp-wave emerges in the continuum
limit. We have also checked that the supersymmetry generators obey the
expected algebra if we limit ourselves to the states that remain light in
the continuum limit and that supersymmetry transformations do not mix the
light states with heavy lattice modes. Our approach crucially depends on the
fact that world-sheet fermions of the IIB string belong to the same
representation of $SO(8)$. It would be interesting to see whether a similar
construction can be made to work for the IIA string where the two sets of
fermions transform differently under $SO(8)$.

We have not discussed multi-string states and  have not addressed the
problem of
constructing interaction vertices. We believe, however, that it should be
relatively straightforward to make the appropriate modifications of the
original string bit model \cite{Verlinde} given our results about fermion
doubling.

One of the main motivations for considering string bit models is the hope
that they may provide an adequate dual description of the SYM theory \cite
{Verlinde,Dhar:2003fi}. In the context of the pp-wave strings, the SYM
operators that are dual to fermion states of the string are known and were
shown to combine into complete supermultiplets \cite{Beisert:2002tn}, even
before taking the continuum limit, which shows that supersymmetry is really
important here. It is also worth mentioning that discretized analogs of the
string Hamiltonian which naturally arise in the SYM theory are Hamiltonians
of integrable spin chains \cite
{Minahan:2002ve,Beisert:2003tq,Belitsky:2003ys}. It would be very
interesting to find connections between the integrable spin chains
and the string bit models.


\subsection*{Acknowledgments}

We are grateful to N.~Beisert, J.~Minahan, C.~Sochichiu, M.~Staudacher and
H.~Verlinde for discussions. This work was supported in part by the Swedish
Research Council (VR). U.D. is a Royal Swedish Academy of Sciences Research
Fellow supported by a grant from the Knut and Alice Wallenberg Foundation.
The work of K.Z. was supported, in part, by RFBR grant 02-02-17260 and grant
00-15-96557 for the promotion of scientific schools.

\end{document}